\documentclass{PoS}

\title{The Baikal-GVD neutrino telescope: First results of multi-messenger studies}

\ShortTitle{The Baikal-GVD first results of multi-messenger studies}

\author{A.D.~Avrorin$^a$, A.V.~Avrorin$^a$, V.M.~Aynutdinov$^a$, R.~Bannash$^g$, I.A~Belolaptikov$^b$, V.B.~Brudanin$^b$, N.M.~Budnev$^c$, G.V.~Domogatsky$^a$, A.A.~Doroshenko$^a$, R.~Dvornick\'y$^{b,h}$, A.N.~Dyachok$^c$, Zh.-A.M.~Dzhilkibaev$^a$, L. Fajt$^{b,h,i}$, S.V~Fialkovsky$^e$, A.R.~Gafarov$^c$, K.V.~Golubkov$^a$, N.S.~Gorshkov$^b$, T.I.~Gress$^c$, R.~Ivanov$^b$, K.G.~Kebkal$^g$, O.G.~Kebkal$^g$, E.V.~Khramov$^b$ , M.M.~Kolbin$^b$, K.V.~Konischev$^b$, A.V.~Korobchenko$^b$, A.P.~Koshechkin$^a$, A.V.~Kozhin$^d$, M.V.~ Kruglov$^b$, M.K.~Kryukov$^a$, V.F.~Kulepov$^e$, M.B.~Milenin$^a$, R.A.~Mirgazov$^c$, V.~Nazari$^b$, \fbox{A.I.~Panfilov$^a$}, D.P.~Petukhov$^a$ E.N.~Pliskovsky$^b$, M.I.~Rozanov$^f$, E.V.~Rjabov$^c$, V.D.~ Rushay$^b$, G.B.~Safronov$^b$, B.A.~Shaybonov$^b$, M.D.~Shelepov$^a$, \speaker{F.~\u{S}imkovic}$^{,b,h,i}$, A.V.~Skurikhin$^d$, A.G.~Solovjev$^b$, M.N.~ Sorokovikov$^b$, I.~\u{S}tekl$^i$, E.O.~Sushenok$^b$, O.V.~Suvorova$^a$, V.A.~Tabolenko$^c$, B.A.~Tarashansky$^c$, and S.A.~Yakovlev$^g$\\
$^a$ Institute for Nuclear Research, Russian Academy of Sciences, Moscow, 117312 Russia\\
$^b$ Joint Institute for Nuclear Research, Dubna, 141980 Russia\\
$^c$ Irkutsk State University, Irkutsk, 664003 Russia\\
$^d$ Institute of Nuclear Physics, Moscow State University, Moscow, 119991 Russia\\
$^e$ Nizhni Novgorod State Technical University, Nizhni Novgorod, 603950 Russia\\
$^f$ St. Petersburg State Marine Technical University, St. Petersburg, 190008 Russia\\
$^g$ EvoLogics Gmbh, Germany\\ 
$^h$ Comenius University, Mlynska Dolina F1, Bratislava, 842 48 Slovakia\\
$^i$ Czech Technical University in Prague, Prague, 128 00 Czech Republic\\
E-mail: \email{fedor.simkovic@fmph.uniba.sk}
}

\abstract{Multi-messenger astronomy is a powerful tool to study the physical processes driving the non-thermal Universe. A combination of observations in cosmic rays, neutrinos, photons of all wavelengths and gravitational waves is expected. The alert system of the Baikal-GVD detector under construction will allow for a fast, on-line reconstruction of neutrino events recorded by the Baikal-GVD telescope and - if predefined conditions are satisfied - for the formation of an alert message to other communities. The preliminary results of searches for high-energy neutrinos in coincidence with GW170817/GRB170817A using the cascade mode of neutrino detection are discussed. Two Baikal-GVD clusters were operating during 2017. The zenith angle of NGC 4993 at the detection time of the GW170817 was 93.3$^{\circ}$. No events spatially coincident with GRB170817A were found. Given the non-detection of neutrino events associated with GW170817, upper limits on the neutrino fluence were established.
}

\FullConference{36th International Cosmic Ray Conference -ICRC2019-\\
		July 24th - August 1st, 2019\\
		Madison, WI, U.S.A.}

\begin{document}

\section{Strategy in search for astrophysical events}
\begin{figure}[!htb]
\includegraphics[width=0.50\textwidth,height=0.35\textwidth]{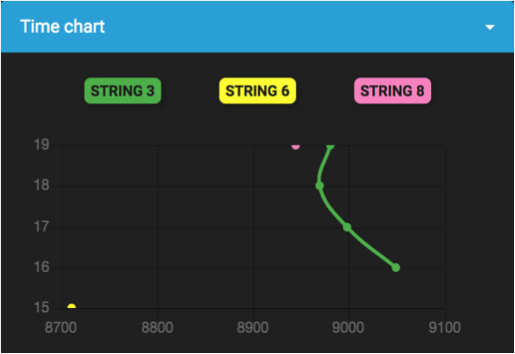}
\includegraphics[width=0.50\textwidth,height=0.35\textwidth]{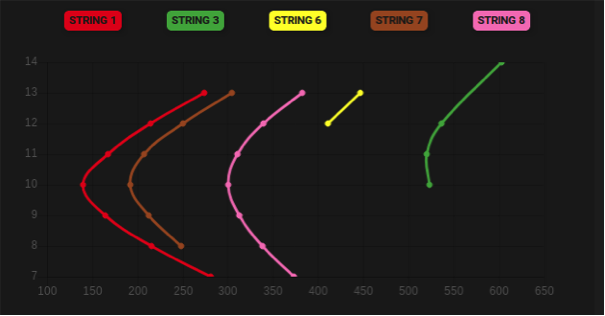}
\caption{Z-coordinates of hit OMs versus arrival time of a Cherenkov photons. Left: Typical reconstruction of a muon track-like event. 
Right: Typical reconstructed cascade-like event.
}
\label{fig-1}
\end{figure}

  We are searching for  correlations in  time and direction with alerts of  other  telescopes at the different received wavelengths i.e. high energy neutrino fluence, electro-magnetic emission, cosmic rays, gravitational waves.  We started with two cosmic events of 2017: GW170817~\cite{GW2017} and IC170922~\cite{TX2017}. Later, according to a signed MoU with the  ANTARES collaboration, we follow up their trigger  of  high energy  neutrinos since December 2018. Two Baikal-GVD clusters were operating during 2017, while in season 2018 there were three operational Baikal-GVD clusters. 
This analysis is based on two modes in the reconstruction of Cherenkov radiation being emitted either by cascade particles or by straight-line moving muons generated in the HE neutrino interactions with Baikal water Ref.~\cite{Cscd},\cite{Trk}. Rates of noise impulses vary between 20 and 100 kHz depending on season and depth, and this  is suppressed by a causality requirement ${\pm}$10ns. Two typical plots for track-like and cascade-like reconstructed events are shown in Fig.~\ref{fig-1}.

In the cascade mode the accuracy of direction reconstruction is about 4.5$^{\circ}$ (median value). The energy resolution averaged by an E$^{-2}$ spectrum of
electron neutrinos is about 30$\%$, about 90\% of these events are within range of 5 TeV$< E <$ 10 PeV. An estimate of the number of 
astrophysical events of an E$^{-2.46}$ spectrum with energies above 100 TeV is about 0.6 per cluster per year and about 0.08 for background events.
Applied quality cuts reject events by number of hits on a string ($>6$), by the value of $\chi^2$ in the reconstruction of the vertex position and by the likelihood value 
in the reconstruction of energy and direction and as well by the product of probabilities of hits and no hits across correspondent optical modules. Two cascade events with $E_{sh} >$ 100 TeV have been selected as astrophycial candidates in the 2015 and 2016 data samples and two new candidates inside the data sample of 2018 (see Ref.~\cite{Cscd}). The values of cuts are slightly varied in dependence on time windows in the follow up analysis, so that they are weaker for shorter times.

In the trajectory mode a procedure based on BDT (in TMVA implementation) was applied. The obtained  angular resolution is less than 1$^{\circ}$, 
while the energy release is not yet estimated. For a total data sample of 2018, only one reconstructed track-event is coming exactly from the same direction as one of the 
alarms, but it is near horizon ($\theta = 110.7^{\circ}$) and does not exceed the expected number of atmospheric muons, also its time is 1.6 hour too early.  
  
\section{Search for neutrinos around GW170817 and IC170922}
\begin{figure}[!htb]
\includegraphics[width=0.45\textwidth,height=0.50\textwidth]{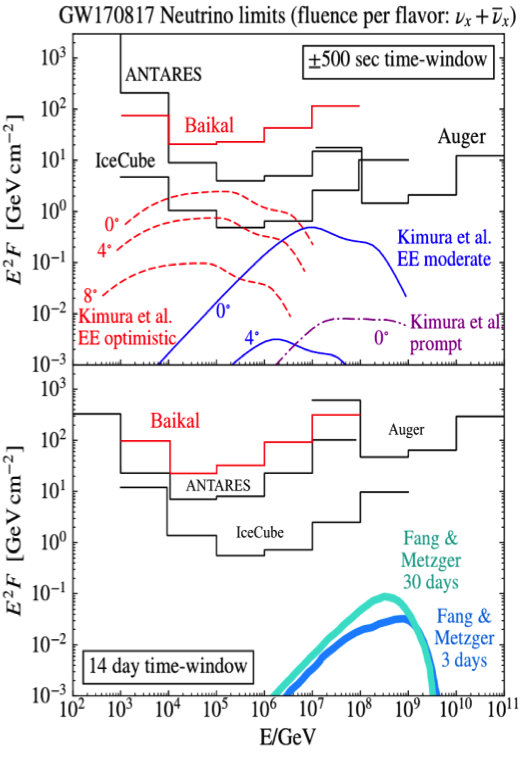}
\includegraphics[width=0.55\textwidth,height=0.35\textwidth]{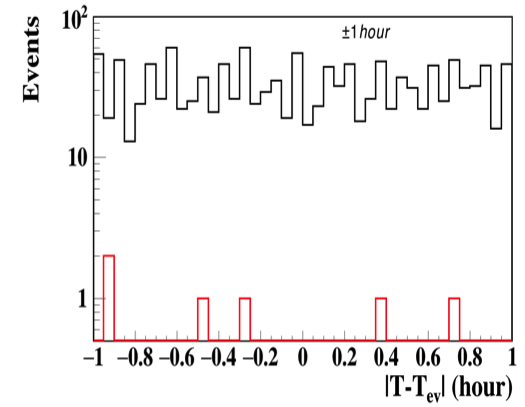}
\caption{Left: Upper limits at 90\% C.L. on the fluence of neutrinos associated with GW170817 for prompt and delayed emission time. 
Right: Distribution of number of events in a time window of ${\pm}$1 hour around IC170922 directed to the blazar TX-0506+056.
}
\label{fig-2}
\end{figure}

Off-line analysis has been done in the cascade mode with two cosmic events over fixed direction and alert time: 
GW170817, gravitational wave origined in galaxy NGC4993 from a merged double neutron star, and IC170922, arising activity of BL LAC galaxy TX-0506+056. 
Two GVD-clusters were operational in 2017. In looking for the fluence in prompt emission ($\pm$500 seconds) from GW170817 the source was located slightly below the horizon for Baikal-GVD  (zenith angle 93$^\circ$), while in the case of IC170922 the source was above the horizon ($\theta$ = 57$^\circ$). No neutrino events associated with these sources have been found in the cascade search mode both in the prompt emissions and in the emission delayed on 14 days after the alert times and restricted by zenith range $74^{\circ}<\theta<150^{\circ}$ at the GVD site. More details of analysis are presented in Ref.~\cite{gvdGW}. Assuming an E$^{-2}$ spectral behavior and equal fluence in neutrino flavors, upper limits at 90\% c.l. have been derived on the neutrino fluence from GW170817 for each energy decade as shown in Fig.~\ref{fig-2} (on left). In following up for the blazar TX-0506+056 the data sample was taken for time window of ${\pm}$1 hour ($\Delta T$ for events with the source time is shown in Fig.~\ref{fig-2} (on right) ) and ${\pm}$1 day inside $45^{\circ}<\theta<126^{\circ}$ as for the GVD site. As result, no neutrino events associated with IC170922A have been recorded
 after all cuts, as seen on angular distance around the source in Fig.~\ref{fig-3} (on left).  

\section{Follow up HE neutrino alerts}

\begin{figure}[!htb]
\includegraphics[width=0.50\textwidth,height=0.35\textwidth]{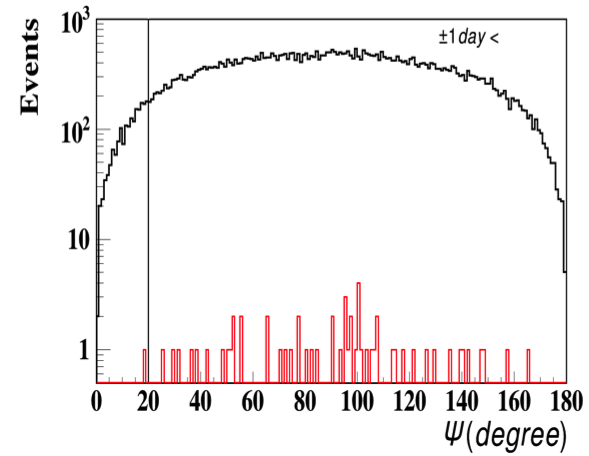}
\includegraphics[width=0.50\textwidth,height=0.38\textwidth]{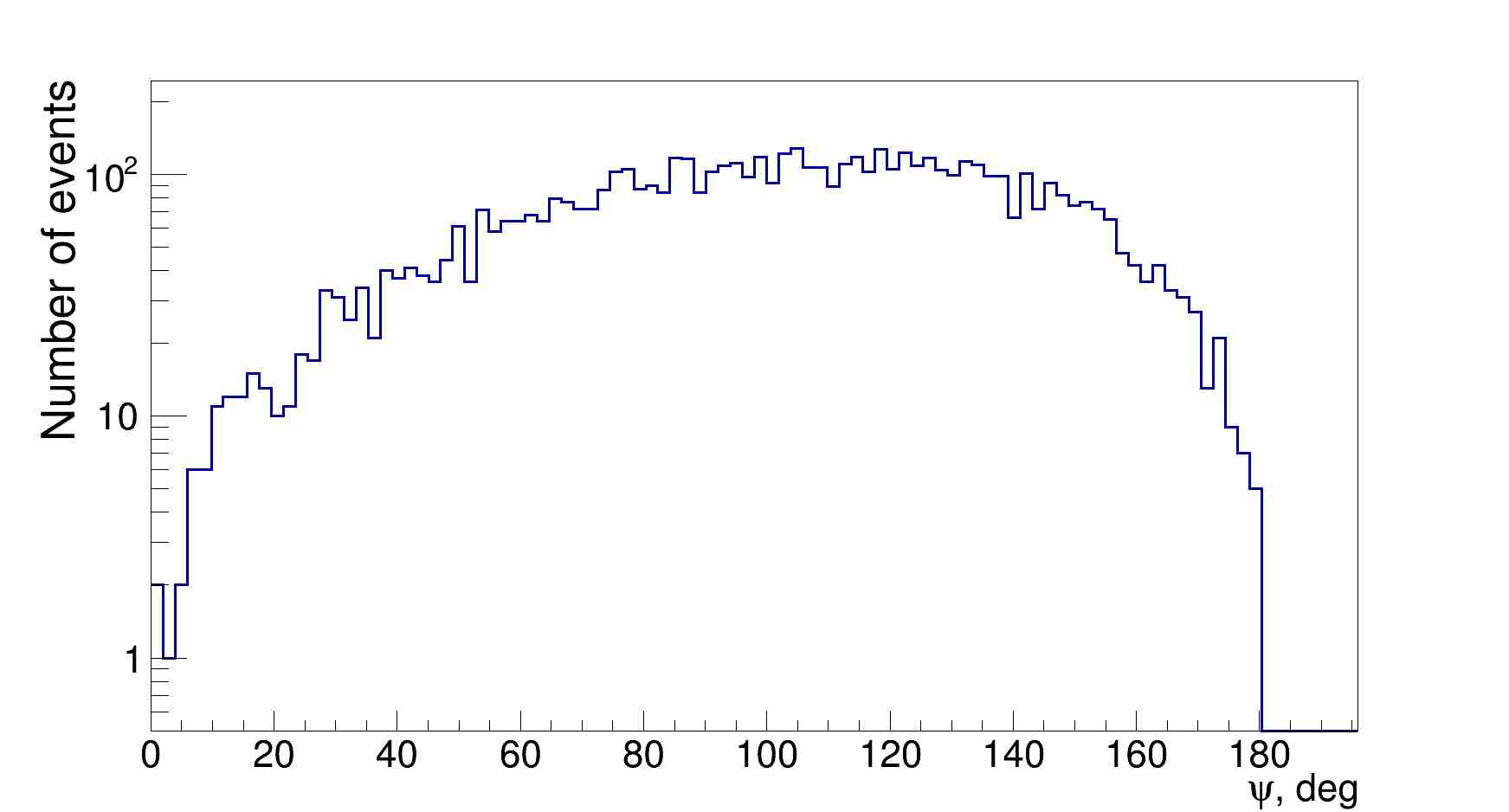}
\caption{Left: Angular distance around the direction of the blazar TX-0506+056. 
Right: Mismatch angle distribution of reconstructed cascades during 6 months observation towards given alert direction.
}
\label{fig-3}
\end{figure}

Following up the ANTARES trigger, we look for events per cluster in time windows of ${\pm}$500 sec, ${\pm}$1 hour and ${\pm}$1 day around alerts inside a half-open cone of 
10$^{\circ}$ and search for coincidence of two or more clusters within 6 $\mu$s for the first ${\pm}$10 sec and in an extended interval of ${\pm}$1 hour around  the trigger. 
As a preliminary result, no time-direction correlations with ANTARES alerts have been found during 6 months of observations.
In best case of happened coincidence within time window of ${\pm}$1day it is shown a distribution of values of mismatch angle with one of the ANTARES sources in Fig.~\ref{fig-3} (right), relatively which there have been found 3 cascade events on 2 clusters among 3 of them inside of half-open cone 5$^{\circ}$. An estimate of atmospheric background has been done on full statistics of reconstructed cascades for 6 months observation towards a given direction and it found to be about 1.75 events for 3 clusters per day for $\psi<10^{\circ}$ and 0.5 events for $\psi<5^{\circ}$. Further off-line analysis will follow.

In summary, the Baikal-GVD design allows to search for HE neutrinos at the early phases of array construction. The GVD developed alert system for multi-messenger studies is in progress.

\section{Acknowledgements}

This work was supported by  the Russian Foundation for Basic Research (Grants 16-29-13032, 17-0201237).

\end{document}